\documentclass[]{JFM-FLM_Au}

\usepackage{amsmath}

\providecommand{\Fr}{\mathrm{Fr}}
\providecommand{\We}{\mathrm{We}}
\providecommand{\Rey}{\mathrm{Re}}

\lefttitle{The Influence of surface tension in hydrodynamics}
\righttitle{ gravity free planar hydraulic jumps}

\title{The influence of surface tension in thin-film hydrodynamics: gravity free planar hydraulic jumps.}

\author{Rajesh K Bhagat\aff{1}}

\affiliation{\aff{1}Department of Applied Mathematics and Theoretical Physics, Wilberforce Road, Cambridge, the UK}

\corresau{rkb29@cam.ac.uk}

\begin{document}
\maketitle

\begin{abstract}
Hydraulic jumps in thin films are traditionally explained through gravity-driven shallow-water theory, with surface tension assumed to play only a secondary role via Laplace pressure. Recent experiments, however, suggest that surface tension can be the primary mechanism. In this work we develop a theoretical framework for \emph{surface-tension–driven hydraulic jumps} in planar thin-film flows. Starting from the full interfacial stress conditions, we show that the deviatoric component of the normal stress enters at leading order and fundamentally alters the balance. A dominant-balance analysis in the zero-gravity limit yields parameter-free governing equations, which admit a similarity solution for the velocity profile. Depth-averaged momentum conservation then reveals a singularity at unit Weber number, interpreted as the criterion for hydraulic control. This singularity is regularised by a non-trivial pressure gradient at the jump. This work establishes the theoretical basis for surface-tension-driven hydraulic jumps, providing analytical predictions for the jump location and structure.

\end{abstract}




\section{Introduction}
Recently, the role of surface tension in hydrodynamics—particularly for kitchen-sink–scale hydraulic jumps—has been debated \cite{bhagat2018origin,bhagat2020experimental,bhagat2020circular,duchesne2022circular}. These jumps, documented since Leonardo da Vinci \citep{marusic2021leonardo}, are classically attributed to gravity. In large-scale open-channel flows, the depth-averaged inviscid (Saint–Venant) equations are hyperbolic, and a hydraulic jump can be interpreted as a shock-like transition connecting supercritical and subcritical states, with criticality associated with the long-wave gravity-wave speed. In the radial setting, however, Bohr \emph{et al.} showed that the ideal (inviscid) shallow-water equations do not allow a determination of the jump position \citep{bohr1993shallow}.

Applying this wave-propagation argument to viscous thin films is theoretically ambiguous. The viscous shallow-water equations used to describe these flows are not hyperbolic; instead, they contain a parabolic contribution arising from wall shear stress, which plays a key role in setting the dynamics \citep{watson1964radial,bohr1993shallow}. Consequently, the system does not admit the same characteristic-based notion of information propagation as in the inviscid theory. Nevertheless, \cite{jannes2012circular} provided experimental evidence supporting the supercritical/subcritical interpretation in a thin film using \emph{silicone} oil. Via Mach-cone measurements, they demonstrated unidirectional propagation of surface disturbances in the fast-moving film and identified the hydraulic jump as a sharp boundary separating super- and subcritical regimes. Physically, this is consistent with viscous stresses decelerating the flow while the film thickens: the long-wave gravity-wave speed increases whereas the mean speed decreases, driving the Froude number toward criticality and leading to choking and the jump. A complementary view, developed by \cite{kurihara1946hydraulic,tani1949water}, attributes the jump to boundary-layer separation driven by an adverse gravitational pressure gradient that builds up as the film thickens.

\begin{figure}
  \centering
  \includegraphics[width=0.9\linewidth]{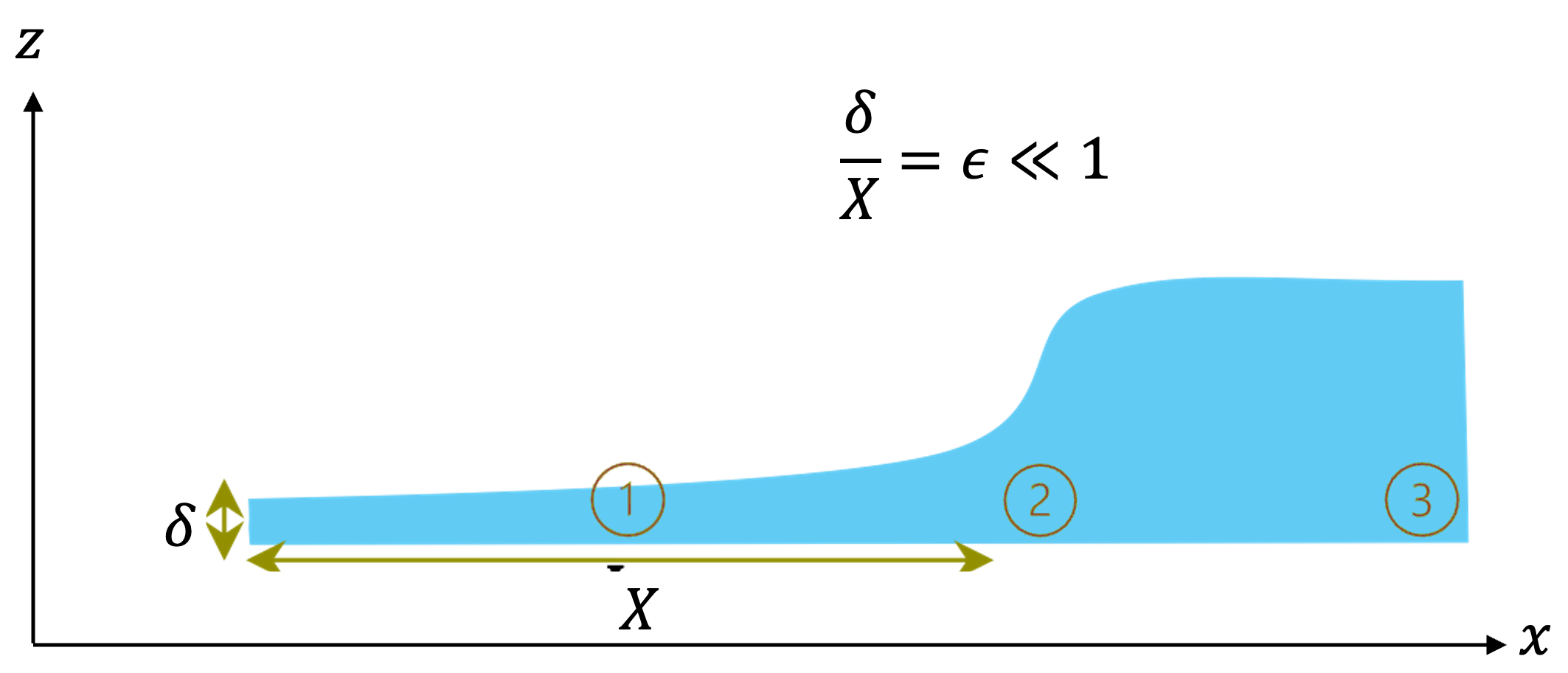}
  \caption{Planar hydraulic‑jump geometry.}
  \label{fig:HJ}
\end{figure}

\cite{bhagat2018origin} challenged the classical picture and proposed a surface-tension–driven mechanism. They argued that hydraulic jumps may be controlled by surface tension, gravity, or a combination of both, and suggested a composite criticality condition based on an energy balance,
\begin{equation}
\We^{-1}+\Fr^{-2}=1,
\label{eq:crit_simple}
\end{equation}
where the local Froude and Weber numbers are defined as
\[
\Fr^{2}=\frac{2\overline{u^{2}}}{gh},
\qquad
\We=\frac{\rho\,\overline{u^{2}}\,h}{\sigma}.
\]
Here $u$ is the velocity, $h$ is the local film thickness, $\rho$ is the density, and $\sigma$ is the surface tension. In thin films they further argued that gravity can be subdominant ($\Fr\gg 1$), so that surface tension provides the dominant control.

While much of the debate has centred on circular jumps, the fundamental balance between surface tension and gravity remains unresolved. To separate physical mechanisms from geometric effects, we consider steady, planar, thin-film channel flow. Under conventional shallow-water scalings—normalising the pressure either by a dynamic scale (kinetic-energy density) or by hydrostatics—the leading-order balances are hydrostatic in the wall-normal direction and viscous--gravity in the streamwise direction \cite{balmforth2004dynamics,dhar2020planar,razis2021continuous}. For nearly flat films, this would suggest that surface tension does not enter at leading order.

We challenge that inference. We show that, beyond the Laplace pressure, the deviatoric normal stress can enter the leading-order balance and, in a zero-gravity thin-film regime, control the dynamics. We first analyse the zero-gravity limit and develop a theoretical framework for surface-tension–driven hydraulic control and jumps, and then validate the predictions against numerical simulations.

The paper is organised as follows. Section~\ref{sec:wave} interprets criticality via a capillary--gravity wave argument. Section~\ref{sec:Theory} states the governing equations and boundary conditions, derives the interfacial conditions without \emph{a priori} assumptions, and scales them to identify the leading-order terms. Section~\ref{sec:CriticalRegime} develops the asymptotics: a dominant-balance analysis identifies the characteristic scales and yields a parameter-free system, from which a similarity solution and a depth-integrated momentum equation are obtained. The latter exhibits a singularity at a unit \emph{effective} local Weber number, which is regularised by pressure.  Section~\ref{sec:conclusions} summarises the findings.

\section{Hydraulic jump as a zero-mode standing wave}
\label{sec:wave}
We frame the steady hydraulic jump as a \emph{wave-pinning} problem: a location
where a surface wave propagating against the flow is held stationary by the
background current. In this sense, the jump may be viewed as a \emph{standing
wave} in the laboratory frame (a \emph{zero-mode}, $\omega=0$) of the
gravity-capillary dispersion relation. Using silicone oil as the working
fluid, \citet{jannes2012circular} performed Mach cone experiments and identified
subcritical and supercritical regions separated by the jump, consistent with a
blocking point where the background flow first matches the relevant long-wave
propagation speed.

For small free-surface disturbances on a layer of depth $h$ with a uniform
current $U$ in the streamwise direction, the Doppler-shifted dispersion relation is
\begin{equation}
(\omega - kU)^2 \;=\; \Big(gk+\frac{\sigma}{\rho}k^3\Big)\tanh(kh),
\qquad \Omega := \omega - kU,
\label{eq:disp}
\end{equation}
where $\omega$ and $k$ are the laboratory-frame frequency and wavenumber, and
$\Omega$ is the intrinsic (fluid-frame) frequency.

\subsection{Gravity-induced jump: setting $\sigma=0$}
In the absence of surface tension and in the shallow-water limit ($kh\ll1$),
\eqref{eq:disp} reduces to $(\omega-kU)^2\simeq gh\,k^2$, i.e.\ a non-dispersive
relation with phase and group speeds equal to $c=\sqrt{gh}$. A steady (zero-mode)
standing wave requires $\omega=0$, hence $\Omega=-kU$ and
$U = |\Omega/k| = c = \sqrt{gh}$, which yields the classical criterion
\begin{equation}
U^2 = gh \quad\Longleftrightarrow\quad \mathrm{Fr}^2 = 1,
\end{equation}
with $\mathrm{Fr}:=U/\sqrt{gh}$.

\subsection{Surface-tension (capillary)-induced jump: setting $g=0$}
The capillary part of \eqref{eq:disp} is dispersive: the intrinsic phase speed
$c_{p0}(k)=\Omega/k$ increases with $k$. In \emph{deep water} ($kh\gg1$) this
scales as $c_{p0}\sim\sqrt{(\sigma/\rho)\,k}$ and therefore becomes unbounded
as $k\to\infty$. If one considers arbitrarily short capillary ripples, this appears
to imply that for any given mean flow $U$ there always exists a sufficiently short
wave that can propagate upstream, so the flow would never be truly ``supercritical''
with respect to \emph{all} capillary modes.

The resolution is that hydraulic control concerns depth-averaged communication
of mass and momentum. Only \emph{long} waves (in the shallow-water sense) have
a substantial depth-integrated signature; very short capillary ripples are
confined to a thin interfacial layer and are strongly damped by viscosity.
Accordingly, we restrict attention to the \emph{long-wave} band
\begin{equation}
 S = \{\,k:\ 0<kh\le \psi\,\},\qquad \psi=O(1),
\end{equation}
where the dimensionless cutoff $\psi$ serves as an \emph{effective} upper limit
for hydraulically relevant waves. Besides the formal shallow-water validity
($kh\lesssim \psi_{\rm SW}$), practical limits arise from (i) \emph{excitation}:
the background inhomogeneity of length $L$ injects little energy at $k\gg 1/L$
(so $\psi_{\rm exc}\sim h/L$), and (ii) \emph{dissipation}: high-$k$ waves are
strongly damped (viscous decay $\gamma\sim 2\nu k^2$), giving an attenuation
length $\ell_{\rm damp}\approx c_g^{\rm lab}/(2\nu k^2)$ that falls rapidly with
$k$. Modes with $\ell_{\rm damp}$ shorter than a few wavelengths do not
contribute to a steady state. We therefore take
\[
\psi=\min\{\psi_{\rm SW},\psi_{\rm exc},\psi_{\rm diss}\}
\]
and treat it as an $O(1)$ constant to be fixed by the flow configuration.
Typically $\psi_{\rm exc}$ and $\psi_{\rm diss}$ exceed unity (see Appendix~1), so
the effective cutoff is often set by the shallow-water limit and we may take
$\psi = \psi_{\rm SW}$.

Isolating capillarity by setting $g=0$, the dispersion relation is
\begin{equation}
\Omega^2=\Big(\tfrac{\sigma}{\rho}k^3\Big)\tanh(kh),
\qquad
c_{p0}(k)=\frac{\Omega}{k}
=\sqrt{\tfrac{\sigma}{\rho}\,k\,\tanh(kh)}.
\label{eq:cap_disp}
\end{equation}
A \emph{zero-mode} (standing wave in the laboratory frame) is obtained by
imposing $\omega=0$, which gives $\Omega=-kU$ and therefore the phase-speed
matching condition
\begin{equation}
U \;=\; c_{p0}(k)
\quad\Longleftrightarrow\quad
U^2 \;=\; \frac{\sigma}{\rho}\;k\,\tanh(kh).
\label{eq:cap_zero_mode}
\end{equation}
Within the admissible long-wave band $S$, $c_{p0}(k)$ monotonically increases with
$k$, so the \emph{last} steady long wave to survive is the \emph{fastest}
admissible long wave at the band edge $k_{\max}=\psi/h$. Evaluating
\eqref{eq:cap_zero_mode} at $k_{\max}$ yields the edge-of-band threshold
\begin{equation}
U_{\mathrm{crit}}^{2}
\;=\;\frac{\sigma}{\rho}\,\frac{\psi}{h}\,\tanh\psi,
\qquad (g=0,\ \text{steady zero-mode}).
\label{eq:Ucrit_cap_exact}
\end{equation}
We define the local \emph{wave-Weber} number based on the band-edge wavenumber
$k_{\max}=\psi/h$,
\begin{equation}
\mathrm{We}:=\frac{\rho U^{2}h}{\sigma\,\psi^{2}},
\label{eq:We_def}
\end{equation}
for which \eqref{eq:Ucrit_cap_exact} becomes
\begin{equation}
\boxed{\;
\mathrm{We}_{\mathrm{crit}}^{(\text{steady})}
=\frac{\tanh\psi}{\psi}
\;}.
\label{eq:We_cap_crit}
\end{equation}
Thus, at $g=0$, a steady capillary-controlled jump occurs when $U$ equals the
phase speed of the fastest admissible long capillary wave ($k=\psi/h$).

\subsection{Gravity-capillary induced jump (completed picture)}
\label{sec:gc_complete}

Combining gravity and surface tension in the \emph{steady} setting, we use the
zero-mode condition $\omega=0$ and the admissible long-wave band $0<kh\le\psi$
to obtain
\begin{equation}
U_{\mathrm{crit}}^{2}
=\Big(\tfrac{gh}{\psi}+\tfrac{\sigma}{\rho}\,\tfrac{\psi}{h}\Big)\tanh\psi
\qquad\Rightarrow\qquad
\boxed{\ \frac{1}{\mathrm{Fr}^{2}}+\frac{1}{\mathrm{We}}
=\frac{\psi}{\tanh\psi}\ \approx 1\ ,}
\label{eq:FrWe_steady_exact}
\end{equation}
where $\mathrm{Fr}:=U/\sqrt{gh}$ and $\mathrm{We}$ is defined in \eqref{eq:We_def}.
Criticality corresponds to satisfying \eqref{eq:FrWe_steady_exact}. Rearranging
\eqref{eq:FrWe_steady_exact} yields
\begin{equation}
\boxed{\ \mathrm{Fr}^{2}\;\approx\;1+\frac{1}{\mathrm{Bo}}\ },
\end{equation}
with the Bond number defined as
\begin{equation}
\mathrm{Bo}:=\frac{\rho g h^{2}}{\sigma\,\psi^{2}}
=\frac{h^{2}}{\psi^{2}l_{c}^{2}},
\qquad l_{c}:=\sqrt{\frac{\sigma}{\rho g}}.
\end{equation}
Thus, for $\mathrm{Bo}\gg1$ (thick films, $h\gg \psi\,l_c$) the jump is
\emph{gravity-controlled} with $\mathrm{Fr}\approx1$; whereas for
$\mathrm{Bo}\ll1$ (thin films) the capillary term sets the threshold, which is
equivalently captured by $\mathrm{We}\approx 1$.

This standing wave viewpoint shows that capillarity can, in principle, set hydraulic control even within an inviscid wave-propagation framework.

\section{Theory}
\label{sec:Theory}
We derived the composite criterion for the thin-film hydraulic jump by interpreting the jump as a steady (`zero-mode') standing wave. We now examine the role of surface tension in such films—a point of recent debate— using momentum equation and quantify the conditions under which capillarity, rather than gravity, sets the control.  

We consider a 2-D, steady, incompressible, Newtonian thin film with constant properties. Gravity is set to zero in this section to expose the mechanism cleanly. Cartesian coordinates are used with $x$ streamwise and $z$ wall-normal; the velocity is $\mathbf{u}=(u,w)$. The volumetric flux per unit span is
\begin{equation}
	Q \;=\; \int_{0}^{h(x)} u(x,z)\,dz.
\end{equation}

The steady incompressible governing equations (in conservative form) are
\begin{align}
 &\partial_x u + \partial_z w = 0, \label{eq:conti_cart}\\
 &u\partial_x(u) + w\partial_z(u) = -\frac{1}{\rho}\,\partial_x p + \nu\big(\partial_{xx}u + \partial_{zz}u\big), \label{eq:xmom_cons}\\
 &\partial_x(u w) + \partial_z(w^2) = -\frac{1}{\rho}\,\partial_z p + \nu\big(\partial_{xx}w + \partial_{zz}w\big), \label{eq:zmom_cons}
\end{align}
where \(\nu\) is the kinematic viscosity.

\subsection{Interface geometry and boundary conditions}
We define the free surface by \(J(x,z)=z-h(x)=0\) \citep{bush2003influence}. The unit normal and tangent to the free surface are
\begin{equation}
 \mathbf{n}=\frac{1}{\sqrt{1+h'^2}}\begin{bmatrix}-h'\\ 1\end{bmatrix},\qquad
 \mathbf{t}=\frac{1}{\sqrt{1+h'^2}}\begin{bmatrix}1\\ h'\end{bmatrix},
 \label{eq:norm_tangent}
\end{equation}
and the curvature is
\[
\kappa=\nabla\!\cdot\!\mathbf{n}=-\frac{h''}{\big(1+h'^2\big)^{3/2}}.
\]
We denote the rate-of-strain tensor by 
\[e_{ij}=\tfrac12(\partial_i u_j+\partial_j u_i)\]
 and the Cauchy stress by 
 \[\boldsymbol{T}=-p\,\boldsymbol{I}+2\mu\,\boldsymbol{e}\]
with the dynamic viscosity $\mu=\rho\nu$. We apply no slip and no penetration condition at the wall (\(z=0\)):
\begin{align}
   \text{at the wall:} \quad(z =0) && u =0; \qquad w =0. \label{bc:wall} 
\end{align}
We assume a clean interface, and negligible gas shear and hence the boundary condition at the free surface \(z = h(x)\). The boundary conditions are, 
\begin{subequations}\label{bc:fs}
\begin{align}
 &\text{kinematic:}  && w \,=\, u_s\,h', \quad \text{where,} \quad u_s(x):=u(x,h), \label{bc:kin} \\
 &\text{zero tangential stress:}  && \mathbf{t}\!\cdot\!\boldsymbol{e}\!\cdot\!\mathbf{n}=0, \label{bc:tang} \\
 &\text{normal stress with surface tension:}  && -p + 2 \mu\,\mathbf{n}\!\cdot\!\boldsymbol{e}\!\cdot\!\mathbf{n} \,=\, -\sigma\,\kappa. \label{bc:norm}
\end{align}
\end{subequations}
For an analytical solution we resolve \eqref{bc:tang} and \eqref{bc:norm} further.
\subsubsection{Zero tangential stress} 
Resolved along \(x-z\) using \eqref{eq:norm_tangent}, \eqref{bc:tang} becomes
\begin{equation}
 \bigg[\frac{1-h'^2}{1+h'^2}\big(\partial_z u+\partial_x w\big)
 + \frac{2h'}{1+h'^2}\big(\partial_z w-\partial_x u\big)\bigg]_{z=h}=0.
 \label{eq:tangential_compact}
\end{equation}
Note that the streamwise axis $x$ is generally \emph{not} aligned with the 
free-surface tangent. The zero--tangential--stress condition applies along the 
surface tangent $\mathbf{t}$, not along $x$. Consequently, its projection onto 
the $x$--direction does \emph{not} enforce $\partial_z u=0$ at $z=h$; a small 
residual streamwise shear is permitted. In the long-wave regime 
($|h'|=O(\epsilon)$), this shear is $O(\epsilon^{2})$ and is negligible at 
leading order. For completeness---and to keep the analysis explicit---we retain 
all terms below and discard the $O(\epsilon^{2})$ contribution only when taking  the leading-order limit. Solving \eqref{eq:tangential_compact} for the streamwise shear at the free surface $\big(\partial_z u+\partial_x w\big)$ gives
\begin{eqnarray}
   \label{eq:tangential_solved}
\left.(\partial_z u+\partial_x w)\right|_{z=h}
= -\,\frac{2h'}{\,1-h'^2\,}\,\Big(\partial_z w-\partial_x u\Big)
\\
\nonumber
= \frac{4h'}{\,1-h'^2\,}\,\partial_x u \quad
\text{(using continuity} \quad \partial_z w=-\partial_x u\text{).} 
\end{eqnarray}

Equations \eqref{eq:tangential_compact}–\eqref{eq:tangential_solved} are \emph{exact} for any slope $|h'|<1$ (the long-wave regime).

\subsubsection{Normal stress condition}
The deviatoric component of normal stress is resolves to
\begin{equation}
 n_i e_{ij} n_j
 =\frac{1}{1+h'^2}\Big[(h')^2\,\partial_x u - h'\big(\partial_z u+\partial_x w\big) + \partial_z w\Big]_{z=h},
 \label{eq:nen_general}
\end{equation}
which, using \eqref{eq:tangential_solved} and continuity, reduces to
\begin{equation}
 \boxed{\;n_i e_{ij} n_j
 = -\,\frac{1+h'^2}{1-h'^2}\,\partial_x u\Big|_{z=h}
 = \frac{1+h'^2}{1-h'^2}\,\partial_z w\Big|_{z=h}.\;}
 \label{eq:nen_compact}
\end{equation}
For nearly flat films (\(|h'|\ll1\)) this simplifies to
\begin{equation}
 \boxed{\;n_i e_{ij} n_j\Big|_{z=h}\to \partial_z w\Big|_{z=h} = -\,\partial_x u\Big|_{z=h}.\;}
 \label{eq:nen_small_slope}
\end{equation}
Thus, even at leading order, the deviatoric component of normal stress contribute to the normal stress balance; only in the inviscid limit \(\mu\to0\) removes this contribution, leaving Laplace pressure only.

\subsection{Scaling}
Here we will first carry out a generic nondimensionalisation and later identify the characteristic scales by a dominant-balance argument. Introduce characteristic scales
\[
x=X\,x^*,\quad z=\delta\,z^*,\quad h=\delta\,h^*,\quad
u=\upsilon\,u^*,\quad w=\frac{\delta \upsilon}{X}\,w^*,
\]
where $\delta$ is the characteristic film thickness, $X$ the streamwise length
scale, and $\upsilon$ the velocity scale, and the aspect ratio 
\[
\epsilon\equiv\frac{\delta}{X}\ll 1.\]
In zero gravity free surface flow, pressure should be naturally be scaled with capillary pressure; with $\kappa\sim\delta/X^2$ we set:
\[
p=\frac{\sigma\,\delta}{X^2}\,p^*.
\]
Define the Reynolds and Weber numbers
\[
Re=\frac{\upsilon\,\delta}{\nu},\qquad
We=\frac{\rho\,\upsilon^2\,\delta}{\sigma}.
\]
{For notational convenience, we now drop the asterisks; all variables below are dimensionless unless stated otherwise.} Dropping asterisks, the nondimensional equations are
\begin{align}
 &\partial_x u + \partial_z w = 0, \label{eq:ND_conti}\\
 &\partial_x(u^2) + \partial_z(u w)
 = -\frac{\epsilon^2}{\We}\,\partial_x p
 + \frac{\epsilon}{\Rey}\,\partial_{xx} u
 + \frac{1}{\epsilon \Rey}\,\partial_{zz} u, \label{eq:NSx_nd}\\
 &\partial_x(u w) + \partial_z(w^2)
 = -\frac{1}{\We}\,\partial_z p
 + \frac{\epsilon}{\Rey}\,\partial_{xx} w
 + \frac{1}{\epsilon \Rey}\,\partial_{zz} w. \label{eq:NSz_nd}
\end{align}
The (dimensionless) flux is
\begin{equation}
 \int_{0}^{h} u\,\mathrm{d}z \,=\, F,\qquad F\equiv Q/(\upsilon\,\delta).
\end{equation}
\subsubsection{Scaled boundary condition}
At the wall, we have \(u =0; \quad w =0\), and the kinematic condition at free surface is \(w=u_s h'\). The resolved tangential condition \eqref{eq:tangential_compact} yields, to \(\mathcal{O}(\epsilon^2)\),
\begin{equation}
 \partial_z u \,=\, -\epsilon^{2}\!\left[\partial_x w
 + \frac{2h'}{1-\epsilon^2 h'^2}\big(\partial_z w - \partial_x u\big)\right]_{z=h},
 \label{bc:tangential_nd}
\end{equation}
so the streamwise free‑surface shear is \(\mathcal{O}(\epsilon^2)\). The normal‑stress condition gives
\begin{equation}
 -\frac{p}{\We} + \frac{2}{\epsilon \Rey}\,\partial_z w \,=\, \frac{h''}{\We} + \mathcal{O}(\epsilon^2)\quad (z=h).
 \label{bc:normal_nd}
\end{equation}
Writing the free-surface conditions in this resolved, scaled form exposes the hierarchy in 
\(\epsilon, Re, We\) and allows direct substitution into depth-integrated balances.
\section{Asymptotic analysis for the critical regime}
\label{sec:CriticalRegime}
\subsection{Critical‑regime scales (capillary control)}
We now determine the characteristic scales $(X,\delta,\upsilon)$ by dominant balances: 1) mass conservation, 2) inertia viscous balance, and 3) capillary control, appropriate to the critical regime:
\begin{align}
 &\text{mass conservation:} && Q \sim \upsilon\,\delta, \label{sc:mass}\\
 &\text{inertia–viscous balance:} && u\,\partial_xu \sim \nu\,\partial_{zz}u \;\Rightarrow\; \upsilon\,\delta^2 \sim \nu\,X, \label{sc:iv}\\
 &\text{capillary control:} && \We=\mathcal{O}(1) \;\Rightarrow\; \rho\,\upsilon^2\,\delta \sim \sigma. \label{sc:we}
\end{align}
Solving \eqref{sc:mass}–\eqref{sc:we} for $(X,\delta,\upsilon)$ yields the unique scales
\begin{align}
 X \,=\, \frac{Q^3\,\rho}{\sigma\,\nu},\qquad
 \delta \,=\, \frac{Q^2\,\rho}{\sigma},\qquad
 \upsilon \,=\, \frac{\sigma}{\rho\,Q}.
\end{align}
Hence, 
    \begin{eqnarray}
        Re=\frac{Q}{\nu}=\epsilon^{-1} \quad \text{so} \quad \epsilon Re = 1, \quad \text{and} \quad  We=1, \quad
        \label{eq:dim_num}
    \end{eqnarray}

\subsection{Parameter‑free leading‑order system}
With \eqref{eq:dim_num} inserted in \eqref{eq:ND_conti}–\eqref{eq:NSz_nd} and retaining the dominant long-wave terms and ignoring \(\mathcal{O}(\epsilon^2)\) terms, the governing equations reduce to
\begin{align}
 &\partial_x u + \partial_z w = 0, \label{eq:pf-cont_sc}\\
 &u\,\partial_x u + w\,\partial_z u = \partial_{zz}u, \label{eq:pf-x_sc}\\
 &\partial_x (uw) + \partial_z (w^2) = -\partial_z p + \partial_{zz}w, \label{eq:pf-z_sc}
\end{align}
the dimensionless flux reduces to 
\begin{equation}
 \int_{0}^{h} u\,\mathrm{d}z \,=\, 1,\qquad F\equiv Q/(\upsilon\,\delta) =1,
\end{equation}
and the boundary conditions simplifies to
\begin{align}
 &u=w=0 && (z=0), \label{eq:pf-bc0}\\
 &w=u_s\,h', && (z=h), \label{eq:pf-bc1}\\
 &\partial_z u=0 && (z=h), \label{eq:pf-bc2}\\
 -&p+2\,\partial_z w=h'' && (z=h). \label{eq:pf-bc3}
\end{align}

\subsection{Similarity solution and depth‑averaged relations}
Guided by the leading order streamwise momentum equation \eqref{eq:pf-x_sc} and the boundary conditions \eqref{eq:pf-bc0} and \eqref{eq:pf-bc1}, we adopt a planar similarity profile,
\begin{equation}
 u(x,z)=u_s(x)\,f(\eta),\qquad \eta=\frac{z}{h(x)},
 \label{eq:ansatz}
\end{equation}
with \(f(0)=0\), \(f(1)=1\), \(f'(1)=0\). The flux constraint gives
\begin{equation}
 \int_{0}^{h} u\,\mathrm{d}z=1\quad\Rightarrow\quad u_s\,h\,I_1=1,\qquad I_1:=\int_0^1 f(\eta)\,\mathrm{d}\eta.
 \label{eq:flux}
\end{equation}
From the flux constraint \eqref{eq:flux} we have 
\(u_sh = \text{constant}.\) Using continuity \eqref{eq:pf-cont_sc} with the ansatz \eqref{eq:ansatz} then yields,
\begin{equation}
 w(x,z)=u_s(x)\,h'(x)\,\eta\,f(\eta).
 \label{eq:w-sim}
\end{equation}
Substituting \eqref{eq:ansatz}–\eqref{eq:w-sim} into the streamwise momentum equation \eqref{eq:pf-x_sc} and using \((u_s h=\frac{1}{I_1}=\text{const})\) yields a separated equation
\begin{equation}
 h^2\,\frac{\mathrm{d}u_s}{\mathrm{d}x} \,=\, \frac{f''(\eta)}{f(\eta)^2} \, :=\, -\frac{3}{2}c^2
 \label{eq:sim_sep}
\end{equation}
Note \(f''(\eta)\leq 0\) as the shear stress is maximum at the plate, and \(f'(1) = 0\), and the right‑hand side is a constant by separation. Equation \eqref{eq:sim_sep} gives the same shape function as obtained by \cite{watson1964radial} for an axisymmetric boundary layer flow, and the shape function can be obtained by setting, 
\begin{eqnarray}
  \frac{f''(\eta)}{f^2(\eta)} = -\frac{3}{2}c^2 \nonumber\\
  \implies f'^2 = c^2(1-f^3)
  \label{eq:shape_fact}
\end{eqnarray}
Equation \eqref{eq:shape_fact} can be integrated to obtain the shape factor. Since \(f' \geq 0\), and \(f(1)=1\), we can get \(c = 1.402\), similarly \(I_1 = 0.615\) are the same as \cite{watson1964radial} due to same shape ODE.
 Using \(u_s h=1/I_1\), differentiating and combining with \eqref{eq:sim_sep} gives a constant free‑surface slope

 \begin{eqnarray}
   u_s h\frac{dh}{dx} = \text{constant} =\frac{3}{2}c^2  \\
   \implies \frac{dh}{dx} = {I_1}\frac{3}{2}c^2 \approx 1.8
    \label{eq:sim_sol2}
\end{eqnarray}
the leading order streamwise momentum equation gives a constant slope for the free surface. 

\subsection{Depth‑integrated wall‑normal momentum}
We integrate \eqref{eq:pf-z_sc} across the film to obtain
\begin{equation}
 \frac{\mathrm{d}}{\mathrm{d}x}\!\int_{0}^{h} u w\,\mathrm{d}z \,=\, p(0) + h'' - 2\,\partial_z w\big|_{z=h} + \big[\partial_z w\big]_{0}^{h}.
 \label{eq:wint_peliminated}
\end{equation}
Substituting equations \eqref{eq:ansatz} and \eqref{eq:w-sim} for velocity, we find
\begin{align}
 &\int_{0}^{h} u w\,\mathrm{d}z \,=\, u_s^2\,h\,h'\,I_2, && I_2:=\int_{0}^{1}\eta f(\eta)^2\,\mathrm{d}\eta\approx0.339,\\
 &\partial_z w\big|_{z=h} \,=\, \frac{u_s h'}{h}, && \partial_z w\big|_{z=0}=0.
\end{align}
Using \(u_s=1/(I_1 h)\) from \eqref{eq:flux} gives
\begin{align}
 \frac{\mathrm{d}}{\mathrm{d}x}\!\int_{0}^{h} u w\,\mathrm{d}z
 &= \frac{I_2}{I_1^2}\,\frac{\mathrm{d}}{\mathrm{d}x}\!\left(\frac{h'}{h}\right)
 = \frac{I_2}{I_1^2}\!\left(\frac{h''}{h}-\frac{h'^2}{h^2}\right),\\
 \partial_z w\big|_{z=h} &= \frac{h'}{I_1 h^2}.
\end{align}
Substituting into \eqref{eq:wint_peliminated} and writing \(P_0:=p(0)\) yields the film‑height ODE
\begin{equation}
 \boxed{\;h'' \,=\, \frac{\displaystyle \frac{I_2}{I_1^2}\,\frac{h'^2}{h^2}
 \, -\, \frac{h'}{I_1 h^2} \, +\, P_0}{\displaystyle \frac{I_2}{I_1^2}\,\frac{1}{h} \, -\, 1}\;}
 \label{eq:jump_noTopPressure}
\end{equation}
Equation \eqref{eq:jump_noTopPressure} shows an  apparent singularity when \(\frac{I_2}{I_1^2}\frac{1}{h} =1\) however, this singularity must be regularised by the wall pressure \(P_0\), a Lagrange multiplier enforcing incompressibility.

We revisit the scaled form of the denominator, which can be written as,
\begin{eqnarray}
    \frac{I_2}{I_1^2}\frac{1}{h^*} \equiv I_2{u_s^*}^2 h^* We \equiv I_2\frac{\rho u_s^2 h}{\sigma}.
    \label{eq:we_f}
\end{eqnarray}
\(I_2\frac{\rho u_s^2 h}{\sigma}\) can also be defined as the local Weber number. Confirming the wave argument; equation \eqref{eq:we_f} shows the shock at the location where the local weber number reaches unity. Hence, we divide the flow that is connected by a jump or a shock into two regimes, 
\begin{eqnarray}
   \frac{I_2}{I_1^2}\frac{1}{{h^*}} =&& \begin{cases}
      >1, & \text{Supercritical}\\
      <1, &\text{Subcritical}
      \end{cases} \label{Eq:regimes}
\end{eqnarray} 
\subsection{Comparison with the standing wave result}
 In equation \eqref{eq:We_def}, we set \(U\) as the average velocity of the film. We compare Equation \eqref{eq:We_def} with \eqref{eq:we_f} to find the value of \(\psi\) to get 

 \begin{eqnarray}
     \psi = \frac{I_1}{\sqrt{I_2}} \approx 1. 
 \end{eqnarray}
\section{Results}


\label{sec:results}

Solving the film-height evolution equation \eqref{eq:jump_noTopPressure} requires knowledge of the pressure field. Our leading-order analysis demonstrated that the streamwise pressure gradient $\partial_x p$ is $\mathcal{O}(\epsilon^2)$; thus, within the slowly varying thin-film region, the pressure may be approximated as constant. However, this assumption holds only in the outer region and breaks down near the jump, where pressure gradients become significant to regularise the flow. Consequently, we distinguish between the outer free-surface flow, where $p \approx \text{const}$, and the inner jump region, where the pressure $P_0$ effectively varies to enforce continuity.

With $\mathcal{H}\equiv h'$, the depth-integrated model reduces to a dynamical system:
\begin{eqnarray}
&&h' = \mathcal{H}, \qquad \mathcal{H}' = \frac{N(h,\mathcal{H})}{D(h)}, \label{eq:phase1} \\
&&N(h,\mathcal{H})=\frac{I_2}{I_1^2}\,\frac{\mathcal{H}^2}{h^2}-\frac{1}{I_1}\,\frac{\mathcal{H}}{h^2}+P_0, \label{eq:phase_num}  \\
&&D(h)=\frac{I_2}{I_1^2}\,\frac{1}{h}-1.
\label{eq:phase_den}
\end{eqnarray}
The phase-space structure can be determined by the nullclines obtained by setting $N(h,\mathcal{H})=0$ and $D(h)=0$. One boundary condition is provided by the similarity slope \eqref{eq:sim_sol2}, while the second depends on the initial film height.

The system \eqref{eq:phase1}--\eqref{eq:phase_den} exhibits a critical singularity when the denominator $D(h)$ vanishes, corresponding to the critical Weber number condition derived in \S 4.1. The appearant singularity gives an impression of a breakdown of the model or a shock. However, in the full viscous system, the pressure is not merely a passive variable but acts as a degree of freedom. Specifically, the wall pressure $P_0$ (and its gradient) functions as the \emph{Lagrange multiplier} that enforces the incompressibility constraint.

We hypothesise that when the flow approaches the singularity ($We \to 1$), the constraint of mass conservation forces the pressure gradient to adjust, thereby regularising the equation. This adjustment allows the flow to negotiate the transition from the supercritical thin film ($We > 1$) to the subcritical downstream layer ($We < 1$) through a smooth, albeit rapid, variation in depth---the hydraulic jump. Thus, while the depth-averaged scaling predicts a singularity at $We=1$, the complete pressure field resolves it, ensuring a continuous physical solution.

\section{Conclusions}
\label{sec:conclusions}

We have presented a theoretical framework for planar hydraulic jumps where surface tension acts as the primary control mechanism. By analysing the gravity-capillary dispersion relation, we showed that even for inviscid fluids, capillarity can in principle set hydraulic control, resulting in the hydraulic jump.

We then analysed the viscous shallow-water equations in the zero-gravity limit and demonstrated that the deviatoric component of normal stress plays a leading-order role in the thin-film momentum balance---a contribution often neglected in classical shallow-water theory.

Our dominant-balance analysis identifies a critical singularity where the local Weber number reaches unity ($We=1$), analogous to the Froude number condition in gravity-driven flows. We have shown that this singularity corresponds to the location where the mean flow velocity matches the speed of the fastest admissible capillary wave (the zero-mode), providing a robust criterion for hydraulic control. Furthermore, our analysis suggests that the apparent singularity in the depth-averaged equations is regularised by the pressure field, which acts as a Lagrange multiplier to enforce continuity across the jump. These results establish that surface tension alone is sufficient to pin a hydraulic jump in thin films, challenging the classical gravity-centric view.




\section*{Acknowledgements}
I would like to express my sincere gratitude to Prof. Paul Linden and Dr. Amir Atufi for their stimulating discussions and suggestions.     \\

\begin{appen}

\section{Viscous cutoff (order-of-magnitude)}\label{appA}

Very short capillary waves are attenuated by viscosity. Balancing the decay rate $\sim \nu k^2$ with the capillary frequency $\omega\sim[(\sigma/\rho)k^3]^{1/2}$ gives
$k_\nu \sim \sigma/(\rho\,\nu^2)$.
For water ($\sigma\approx 0.072\ \mathrm{N\,m^{-1}}$, $\rho\approx 10^3\ \mathrm{kg\,m^{-3}}$, $\nu\approx 10^{-6}\ \mathrm{m^2\,s^{-1}}$) this yields $k_\nu \sim 7\times 10^{7}\ \mathrm{m^{-1}}$ ($\lambda\sim 90\ \mathrm{nm}$), far beyond the long-wave range and irrelevant to the control criterion.

\end{appen}\clearpage

\bibliographystyle{jfm}
\bibliography{jfm}



\end{document}